\documentclass{moriond}

\pdfoutput=1

\def\mco{\multicolumn}

\def\be{\begin{equation}}
\def\ee{\end{equation}}
\def\bea{\begin{eqnarray}}
\def\eea{\end{eqnarray}}

\newcommand{\Photo}{}

\begin{document}
\vspace*{4cm}
\title{NIKA2 Performance and Cosmology Program with Galaxy Clusters}

\author{L.~Perotto$^{1}$, R.~Adam$^{2}$, P.~Ade$^{3}$, P.~Andr\'e$^{4}$, M.~Arnaud$^{4}$, H.~Aussel$^{4}$, I.~Bartalucci$^{4}$, A.~Beelen$^{5}$, A.~Beno\^it$^{6}$, A.~Bideaud$^{6}$, O.~Bourrion$^{1}$, M.~Calvo$^{6}$, A.~Catalano$^{1}$, B.~Comis$^{1}$, M.~de Petris$^{7}$,  F.-X.~D\'esert$^{8}$, S.~Doyle$^{3}$, E.~F.~C.~Driessen$^{9}$, A.~Gomez$^{16}$, J.~Goupy$^{6}$, C.~Kramer$^{9}$, G.~Lagache$^{10}$, S.~Leclercq$^{11}$, J.-F.~Lestrade$^{12}$, J.F.~Mac\'ias-P\'erez$^{1}$, P.~Mauskopf$^{3,\,  13}$, F.~Mayet$^{1}$, A.~Monfardini$^{6}$, G.~Pisano$^{3}$, E.~Pointecouteau$^{14}$, N.~Ponthieu$^{8}$, G.~W.~Pratt$^{4}$, V.~Rev\'eret$^{4}$, A.~Ritacco$^{9}$,  C.~Romero$^{11}$,  H.~Roussel$^{15}$, F.~Ruppin$^{1}$, K.~Schuster$^{11}$, A.~Sievers$^{9}$, C.~Tucker$^{3}$, R.~Zylka$^{11}$} 

\address{$^{1}$Univ. Grenoble Alpes, CNRS, Grenoble INP, LPSC-IN2P3, 53, avenue des Martyrs, F-38000 Grenoble, France\\
  $^{2}$Centro de Estudios de F\'isica del Cosmos de Arag\'on (CEFCA), Plaza San Juan, 1, planta 2, E-44001, Teruel, Spain\\
  $^{3}$Astronomy Instrumentation Group, University of Cardiff, UK\\
  $^{4}$AIM, CEA, CNRS, Universit\'e Paris-Saclay, Universit\'e Paris Diderot, Sorbonne Paris Cit\'e, F-91191 Gif-sur-Yvette, France
  $^{5}$Institut d'Astrophysique Spatiale (IAS), CNRS and Universit\'e Paris Sud, Orsay, France \\
  $^{6}$Institut N\'eel, CNRS and Universit\'e Grenoble Alpes, France\\
  $^{7}$Dipartimento di Fisica, Sapienza Universit\`a di Roma, Piazzale Aldo Moro 5, I-00185 Roma, Italy\\
  $^{8}$Univ. Grenoble Alpes, CNRS, IPAG, F-38000 Grenoble, France\\
  $^{9}$Institut de RadioAstronomie Millim\'etrique (IRAM), Granada, Spain\\
  $^{10}$Aix Marseille Universit\'e, CNRS, LAM (Laboratoire d'Astrophysique de Marseille) UMR 7326, 13388, Marseille, France\\
  $^{11}$Institut de RadioAstronomie Millim\'etrique (IRAM), Grenoble, France\\
  $^{12}$LERMA, Observatoire de Paris, PSL Research University, CNRS, Sorbonne Universit\'es, UPMC Univ. Paris 06, F-75014, Paris, France\\
  $^{13}$School of Earth and Space Exploration and Department of Physics, Arizona State University, Tempe, AZ 85287\\
  $^{14}$Universit\'e de Toulouse, UPS-OMP, Institut de Recherche en Astrophysique et Plan\'etologie (IRAP), Toulouse, France\\
  $^{15}$Institut d'Astrophysique de Paris, CNRS (UMR7095), 98 bis boulevard Arago, F-75014, Paris, France\\ 
  $^{16}$Centro de Astrobiolog\'ia (CSIC-INTA), Torrej\'on de Ardoz, 28850 Madrid, Spain}
 
\maketitle

\abstracts{NIKA2 is a dual-band millimetric camera of thousands of Kinetic Inductance Detectors (KID) installed at the IRAM 30-meter telescope in the Spanish Sierra Nevada. The instrument commissioning  was completed in September 2017, and NIKA2 is now open to the scientific community and will operate for the next decade. NIKA2 has well-adapted instrumental design and performance to produce high-resolution maps of the thermal Sunyaev-Zel'dovich (SZ) effect toward intermediate and high redshift galaxy clusters. Moreover, it benefits from a guaranteed time large program dedicated to mapping a representative sample of galaxy clusters via SZ and that includes X-ray follow-ups. The main expected outputs of the SZ large program are the constraints on the redshift evolution of the pressure profile and the mass-observable relation. The first SZ mapping of a galaxy cluster with NIKA2 was produced, as part of the SZ large program. We found a sizable impact of the intracluster medium dynamics on the integrated SZ observables. This shows NIKA2 capabilities for the precise characterisation of the mass-observable relation that is required for accurate cosmology with galaxy clusters.}

\clearpage

%
%

\section{Cosmological motivations for high-resolution observation of high-$z$ clusters}

Galaxy clusters represent a powerful probe for cosmology. As the largest gravitationally bound objects that the hierarchical structure growth have formed, they enclose information on initial conditions, contents and expansion history of the Universe. Their number as a function of mass and redshift and their spatial distribution are used in complement of other observables of the primordial Universe to constrain the parameters of the standard cosmological model, in which the present-day density is dominated by a cosmological constant $\Lambda$ and the cold dark matter (CDM), as well as any extensions to this base flat $\Lambda$CDM model that impact the growth of structures.

Galaxy clusters are efficiently detected via the thermal Sunyaev-Zel'dovich (SZ) effect up to high-redshifts. The thermal SZ effect~\cite{Sunyaev:1972}$^,$~\cite{Sunyaev:1980vz}$^,$~\cite{Birkinshaw:1998qp} is the inverse Compton scattering of the CMB photons on the hot dense ionised gas, the Intra-Cluster Medium (ICM), which accounts for most of the baryonic matter content of the cluster. It has a unique signature on the CMB black-body spectrum, decreasing the CMB intensity at frequencies below $217$~GHz at the benefit of the higher frequencies. As a spectral distortion, the SZ effect is independent of the redshift and thus well-suited for observation of high-redshift clusters. The amplitude of the effect, the Compton parameter $y$, directly depends on the line-of-sight integral of the electronic pressure within the cluster. The Compton parameter is thus a tracer of the pressure within the ICM, which in turn depends on the total mass of the cluster. The relation between the SZ observable and the total mass needs to be calibrated. This can be achieved using total mass measurements obtained from  X-ray observations (assuming clusters are in hydrostatic equilibrium), galaxy velocities, caustics, or from weak lensing. 

CMB experiments such as \emph{Planck}~\cite{Ade:2013skr}$^,$~\cite{Ade:2015gva}, the South Pole Telescope~\cite{Bleem:2014iim} (SPT) or the Atacama Cosmology Telescope~\cite{Hasselfield:2013wf} (ACT) have recently delivered the community with catalogs of about 2000 clusters detected via the thermal SZ effect. In parallel, strong experimental efforts are deployed to reach higher angular resolution for the cluster observation, NIKA2~\cite{Calvo16}$^,$~\cite{NIKA2-Adam}, MUSTANG2~\cite{Mustang2_instru}, ALMA~\cite{alma_sz} amongst others. Cosmological constraints have been derived from the SZ-detected cluster number counts by \emph{Planck}~\cite{Ade:2013lmv}$^,$~\cite{Ade:2015fva}, SPT~\cite{Bleem:2014iim} and ACT~\cite{Hasselfield:2013wf}, as well as from the power spectrum and higher-order statistics within the \emph{Planck} all-sky map of Comptom parameter~\cite{Planckymap}$^,$~\cite{PlancktSZspec}$^,$~\cite{Bolliet2018}. Notably, the \emph{Planck} Collaboration reported a mild tension between the cosmological parameters measured from the CMB primary anisotropies and from the SZ-selected clusters, which is also found using other SZ experiment data (SPT, ACT) or other cluster probes. This can be an evidence of a new physical effect that impacts the growth of structures (e.g. non-minimal neutrino masses, non-unity dark energy equation of state, modified gravity), or this can be an impact of the complex cluster physics on the total mass estimation (deviation from self-similar scenario, non-thermal pressure, redshift evolution, etc.). In particular, in \emph{Planck} study, the pressure profile and mass-observable scaling relation have been calibrated from tSZ and X-rays low-redshift clusters, whereas a redshift evolution effect is not excluded. Thus, accurate cosmology with galaxy clusters and tests for deviations from the standard ($\Lambda$CDM) cosmological model demand a more robust calibration of the mass-observable relation against the redshift and the cluster internal matter distribution. To that aim, high angular resolution observations of high redshift clusters are required. This is one of the main objectives of the NIKA2 experiment.\\




%
%
\section{NIKA2 Instrument and Performance}
\label{Sect:perf}

NIKA2 is a ground-based millimetric experiment installed at the 30-meter telescope of the \emph{Institut de RadioAstronomie Millim\'etrique} (IRAM), situated at Pico Veleta (2800 meters) in the Spanish Sierra Nevada. The NIKA2 Collaboration involves 150 people in 18 institutes (the NIKA2 consortium$\&$IRAM), with scientific objectives that cover large domains of astrophysics and cosmology.    

\subsection{Instrument}

\begin{figure}
  \centering
  \includegraphics[width=0.53\textwidth]{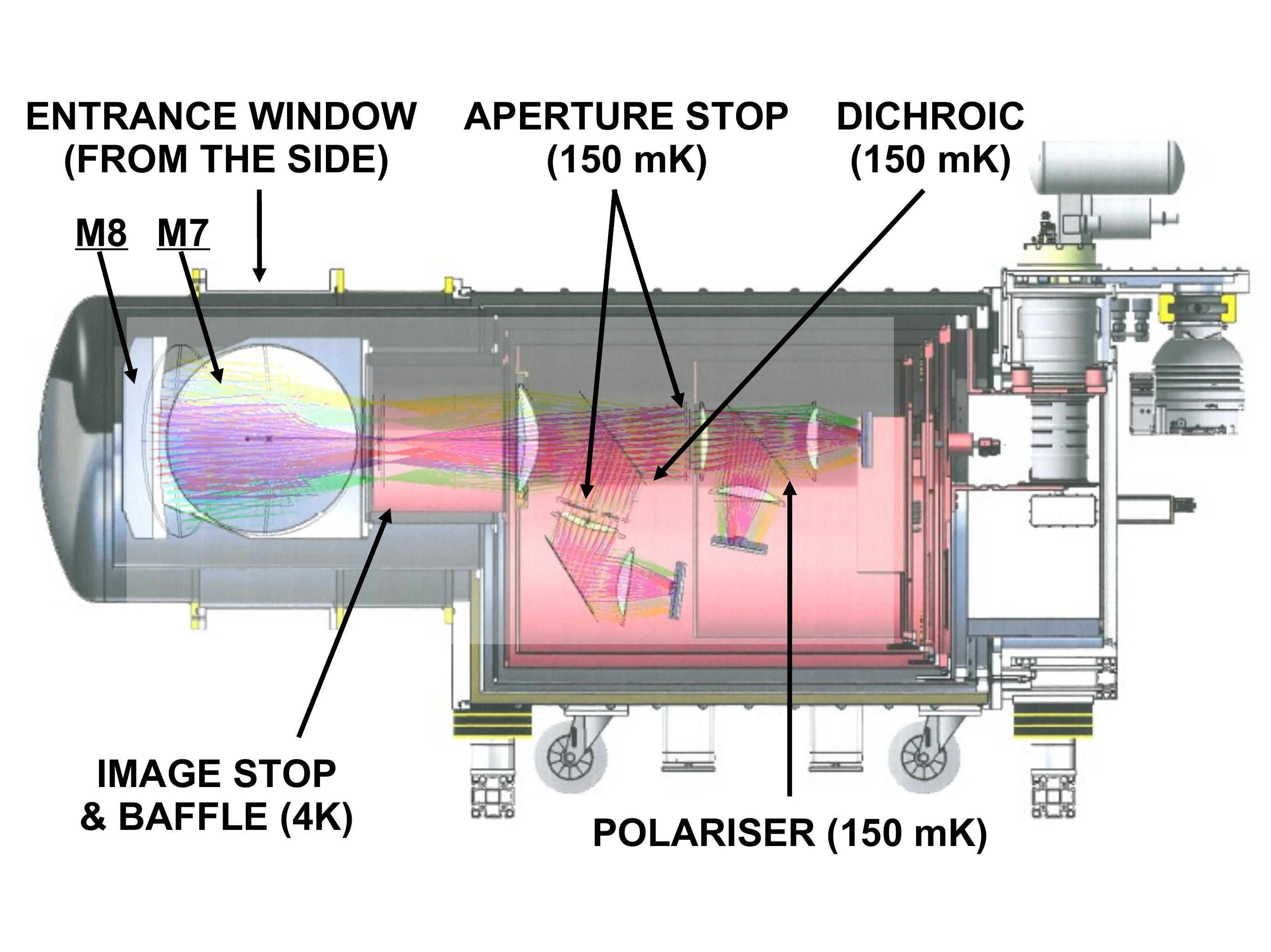}
  \includegraphics[trim=2cm 14cm 4cm 4cm, clip=true,width=0.45\textwidth]{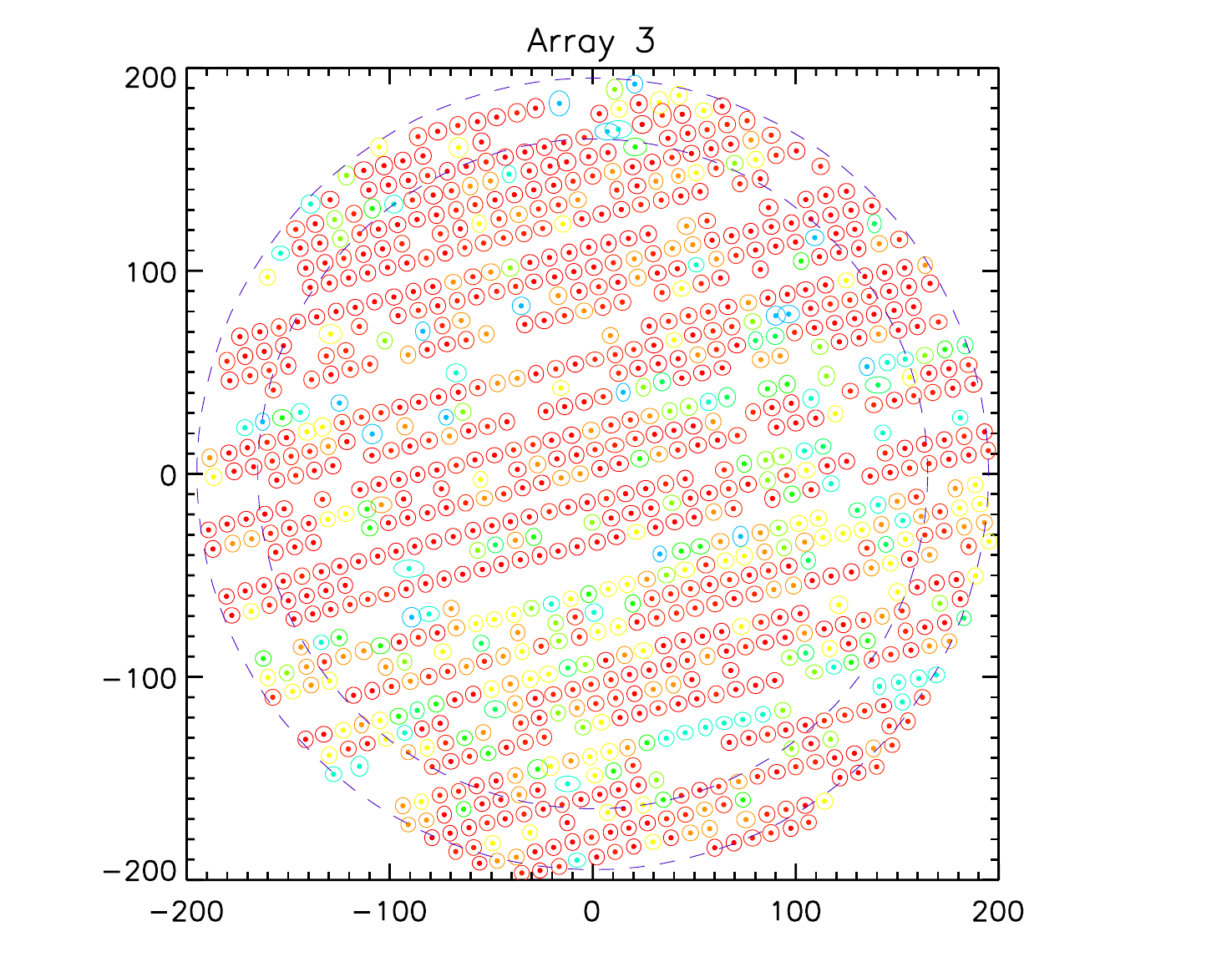}
  \caption{Left: Cross-section of the NIKA2 instrument illustrating the cold optics. Right: Footprint of the detector positions in the field of view for one of the 260~GHz arrays, colour-coded as a function of the number of times they met the selection criteria (from red = valid for all selections to green = valid in one selection). These figures are extracted from Adam et al. (2018)$^{9}$. }
  \label{fig:instru}
\end{figure}

The NIKA2 instrument consists of a multi-thousand supraconductor detector (Kinetic Inductance Detectors) camera, observing in two frequency bands at 150 and 260 GHz with an angular resolution below 20 arcsec and a large (6.5 arcmin) field of view. After a first light in October 2015, it underwent an extensive commissioning phase including an upgrade in September 2016 and a science verification phase in April 2017. NIKA2 is now open for science operation since October 2017 and will be a resident instrument at the 30-meter telescope of IRAM for the next decade. Its novel technology and instrumental design were sucessfully tested with the NIKA experiment~\cite{Monfardini11}$^,$~\cite{Bourrion12}$^,$~\cite{Calvo13}, which is a NIKA2 pathfinder down-scaled by about a factor ten, operated at the IRAM 30-m telescope as an open instrument during the winter 2014.   

The light collected by the 30-m telescope primary mirror reaches NIKA2 focal plane through a dedicated warm optics (inside the receiver cabin), which comprizes four mirrors, and a dedicated cold optics (inside the cryostat), which is illustrated in Fig.~\ref{fig:instru}. The light enters the side of the cryostat (about 30~K), is directed toward the coldest part by M7 and M8 mirrors. Within the 150~mK stage, a dichroic splits the beam in two frequency bands, the 150-GHz band in reflection and the 260-GHz band in transmission. The flatness of this 30-cm diameter dichroic submitted to extreme temperature is ensured by the used of a novel air-gap filled technology and a reinforced support. The 260~GHz beam is further splits in two polarisation components by a grid polariser.

At 150 and 260~GHz, angular resolutions below $20~\rm{arcsec}$ can be reached given the 30-meter telescope aperture. In order to fully exploit the 30-meter telescope capabilities, that is filling the $6.5~\rm{arcmin}$ field of view without degrading the angular resolution, arrays of thousands detectors are needed at the focal plane. For NIKA2, the choice has been made to use Kinetic Inductance Detectors (KIDs) which offers the required multiplexing capabilities.  KIDs are superconducting RLC resonators, operated at 150~mK, well-below their critical temperature. Incoming radiation changes the kinetic-component of the inductance (by Cooper pair breaking), so that the signal varies proportionally with the resonance frequency. Raw data are thus variations of the resonance frequency of each KID. Moreover, KIDs are naturally multiplexable in frequency: they can be connected to a unique feed-line that carries up to 200 frequency tones. NIKA2 comprizes two arrays of 1140 KIDs connected by eight feed-lines in the 260~GHz channel and an array of 616 KIDs connected by four feed-lines at 150~GHz~\cite{Calvo16}$^,$~\cite{NIKA2-Adam}. 
Each feed-line is monitored a dedicated readout electronic board, as part of the NIKEL electronics~\cite{Bourrion12}. 

\subsection{Calibration and Performance}

\begin{table*}[t]
  \centering
  \caption[]{Summary of the principle characteristics and performance of the NIKA2 instrument. \label{sumperf}}
  \begin{tabular}{|c|c|c|c|c|}
    \hline
    & \multicolumn{3}{|c|}{ }  & \\
    Channel & \multicolumn{3}{|c|}{260 GHz} & 150 GHz \\
    & \multicolumn{3}{|c|}{1.15 mm}     &  2 mm \\ 
    \hline
    Arrays & A1 & A3  & A1\&3 & A2 \\
    \hline
    Number of designed detectors       & 1140      &  1140    &    &    616      \\
    Number of valid detectors $^{1}$     &  952      &   961    &   &    553      \\ 
    \hline
    FOV diameter [arcmin]     &   6.5              &  6.5              &   6.5        &    6.5        \\
    FWHM [arcsec]             &   $11.3 \pm 0.2$   &  $11.2 \pm 0.2$  &   $11.2 \pm 0.1$           &  $17.7 \pm 0.1$ \\      
    Beam efficiency $^{2}$ [\% ]   & $55 \pm 5$  &  $53 \pm 5$  &  $60 \pm 6$        &     $75 \pm 5$ \\
    \hline 
    rms calibration error [\%]            & \multicolumn{3}{|c|}{7.5}  & 3  \\
    \hline
    Model absolute calibration uncertainty [\%] &  \multicolumn{4}{|c|}{5} \\
    \hline
    NEFD $^{3}$  [mJy.s$^{1/2}$]        &    &     & 33$\pm$2      & 8$\pm$1  \\
    Mapping speed $^{4}$ [arcmin$^2$/h/mJy$^2$]  &   &   & 67 - 78  & 1288 - 1440 \\
    & & & & \\
    \hline
    \mco{5}{l}{ \begin{minipage}{6in} \small{$(1)$ Number of detectors that are valid at least for two different beam map scans. $(2)$ Ratio between the main beam power and the total beam power up to a radius of 250$^{\prime \prime}$. $(3)$ Average NEFD during the February 2017 observation campaign, extrapolated at $\tau = 0$. $(4)$ Average mapping speed during the February 2017 observation campaign, extrapolated at $\tau = 0$.} \end{minipage} } \\
    
  \end{tabular}
\end{table*}

The first step of any KID-based data analysis consists in the field-of-view (FoV) reconstruction that is, matching the KID frequency tones to positions on the sky. We use deep integration scans of about 20 minutes toward bright point sources, the so-called beam-map scans, to perform individual maps per KID. From these maps, we derive i) KID positions on the FoV, ii) beam properties, iii) detector inter-calibration. Beam-map scans are also used to perform a KID selection from a series of quality criteria. Although all the (2,900) built KID are responsive, some of them are affected by cross-talk or their frequency tuning can be lost during a given scan. In the right panel of Fig.~\ref{fig:instru}, we present the KID positions in the FoV colour-coded as a function of the number of times they met the selection criteria. We find a fraction of valid KIDs, which is defined as the selected detectors in at least two beam-map scans, of $84\%$ at 260~GHz and $90\%$ at 150~GHz.

In combining the signal of all selected detectors of an array, we probe the full beam pattern. The main beam is well-modeled by a 2D Gaussian of FWHM of $11.2 \pm 0.1$~arcsec at 260~GHz and of $17.7 \pm 0.1$ at 150~GHz. The beam efficiency, which is evaluated as the ratio of the main beam to the error beam solid angle integrated up to 250 arcmin, is $55\%$ at 260~GHz and $75\%$ at 150~GHz, in agreement with expectations for an instrument without any dedicated optical coupling (no feed horns).

Regarding the noise properties, the dominant noise component is the atmospheric fluctuations, which induce strong 1/f noise spectrum. However, this noise is seen by all the detectors while the signal depends on the detector positions in the FoV. Thus atmospheric noise can be decorrelated. After decorrelation, sub-dominant correlated noise residuals from the atmosphere and the electronics are left in the maps, but do not affect the noise scaling down with integration time: we checked that the flux uncertainties reduce as expected as the inverse of the square root of the integration time.

The absolute calibration relies on flux density expectation for planets~\cite{Bendo2013}. We check the stability of the photometry using secondary calibrators monitored at Plateau de Bure. We evaluate our calibration uncertainties as the relative rms of the measured-to-expected flux ratio for secondary calibrators including data from three observation campaigns. We found uncertainties of $8\%$ at 260 GHz and $3\%$ at 150 GHz, which improve or match the level of accuracy of other recent ground-based expirements observing in comparable frequency bands (such as SCUBA2~\cite{Dempsey2013}).  

We primary assess the sensitivity by evaluating the noise equivalent flux density (NEFD), defined as the flux uncertainties obtained at the center of the FoV after one second of integration. This quantity is needed to predict the required observation time to reach a given SNR at the center of the map. We find NEFD extrapolated at zero atmospheric opacity of $33 \pm 2\, \rm{mJy} \cdot \sqrt{s}$ at 260 GHz and $8 \pm 1\, \rm{mJy} \cdot \sqrt{s}$ at 150 GHz. However, we also evaluate the mapping speed, defined as the sky area that is covered in one hour of observation with a SNR of unity, which is a better estimator of the mapping capabilities. We find mapping speeds of $75 \pm 5 \, \rm{arcmin}^2/h/\rm{mJy}^2$ at 260 GHz and $1350 \pm 75 \, \rm{arcmin}^2/h/\rm{mJy}^2$ at 150 GHz. Moreover, we directly verified using observation of Pluto that mJy sources can be detected in less than one hour of integration time. The main characteristics and performance of NIKA2 are summarized in Table 1.   

NIKA2 has thus demonstrated high sensitivity and state-of-the art mapping capabilities of the diffuse emission. It has two observation channels, the 150 GHz band, for which the thermal SZ is negative, and the 260 GHz band, for which the SZ signal is slightly positive, and a large 6.5~arcmin FoV enclosing the \emph{Planck} satellite beam, while its angular resolution is 17 times better. NIKA2 has thus well-adapted instrumental design and performance for high-resolution SZ observation of intermediate and high redshift clusters.

%
%
\section{NIKA2 Cosmology program with galaxy clusters and First results}

\begin{figure}
  \centering
  \includegraphics[width=0.49\textwidth]{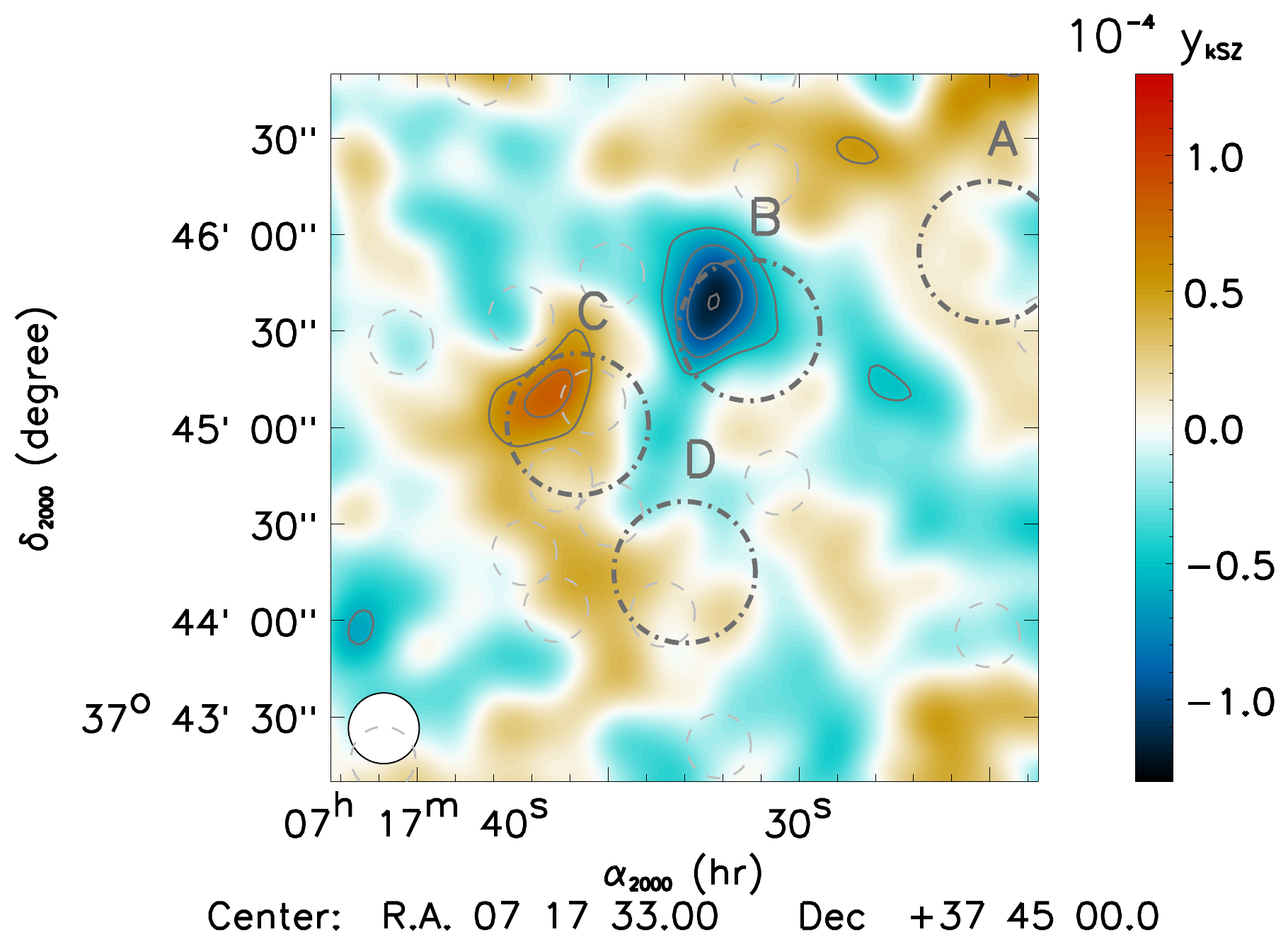}
  \hspace{4mm}
  \includegraphics[width=0.44\textwidth, clip=true, trim=0cm -0.7cm 0cm 0cm]{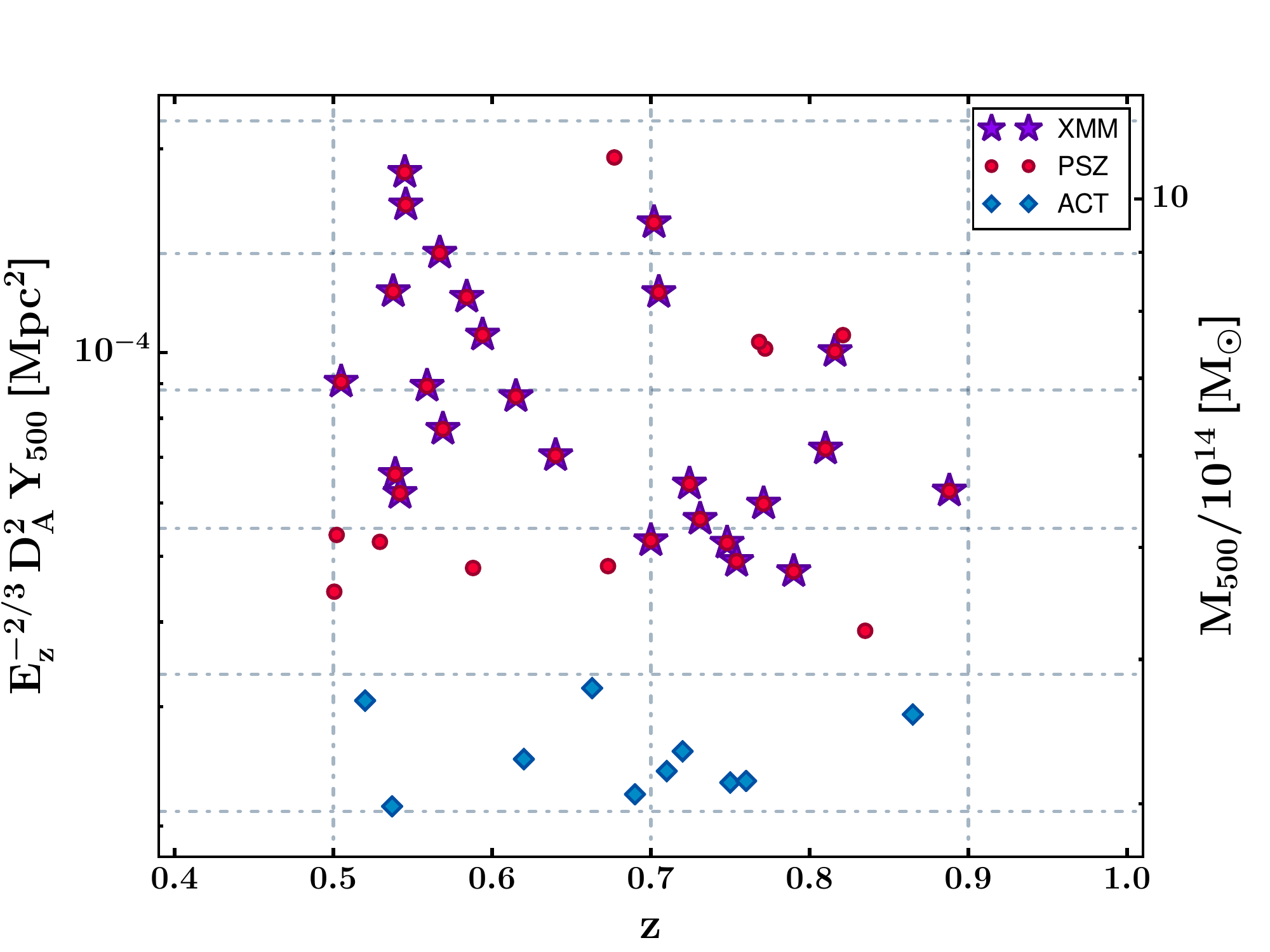}
  \caption{Left: Map of the kinetic SZ effect toward \mbox{MACS~J0717.5+3745} using NIKA pathfinder data, as discussed in Adam et al. (2017)$^{31}$. Right: NIKA2 Guaranteed-time cosmology program sample of galaxy clusters}
  \label{fig:nikanika2}
\end{figure}

\subsection{NIKA Pathfinder SZ pilot studies}

High-resolution SZ mapping capabilities using KID-based experiments have been demonstrated with NIKA, the NIKA2 pathfinder, through a series of pilot studies. The first SZ observation ever obtained with KIDs is the NIKA SZ map of the galaxy cluster RX J1347.5-1145, a well-known massive intermediate redshift galaxy cluster~\cite{Adam:2013ufa}. NIKA also paves the way for NIKA2 SZ analysis, by proposing several novel analysis methods: a multi-probe approach combining X-ray and SZ observations for high-redshift cluster thermodynamics~\cite{Adam:2014wxa}, point source removal~\cite{Adam:2015bba}, a non-parametric method to measure the pressure profile~\cite{Ruppin:2016rnt}$^,$~\cite{Romero:2017xri}, a detection method of substructures in clusters~\cite{Adamsubstruct}. Moreover, breakthrough results have been obtained, including the first map of the gas temperature in galaxy clusters using  SZ~\cite{Adam:2017mlj} and the first map of the kinetic SZ in a galaxy cluster~\cite{Adam:2016abn}.

The kinetic SZ, which is due to the bulk velocity of the ICM, has a spectral signature on the CMB spectrum that differs from the thermal SZ one. By combining maps observed in different channels, both SZ components can be in principle separated. This has been achieved for the first time by combining NIKA 150 GHz and 260 GHz maps toward a multiple merger cluster the inner substructures of which reach extreme velocities, that is the well-known MACS J0717.5+3745. The NIKA kinetic SZ map toward MACS J0717.5+3745 at $z= 0.55$ is presented in the left panel of Fig.~\ref{fig:nikanika2}. The data set comprizes about 13h of observation using NIKA during Winter 2015 in good weather conditions. This kSZ map was used in combination to X-ray observation from XMM-Newton to derive the first resolved map of the gas velocity under the assumption of a gas model~\cite{Adam:2017mlj}. 
 
\subsection{NIKA2 Guaranteed time SZ Large Program}

Cosmology with galaxy clusters observed via SZ is a major scientific goal of NIKA2. It is supported by the guaranteed time SZ Large program (LP-SZ), dedicated to the high-resolution mapping of a large sample of high-redshift clusters. The LP-SZ targets a representative sample of 50 galaxy clusters at intermediate and high-redshifts ($0.5 \le z \le 0.9$), selected from the \emph{Planck} and ACT cluster catalogs and complemented with X-ray follow-up using XMM-Newton and Chandra. The distribution of the LP-SZ sample as a fonction of the mass and the redshift is presented in the right panel of Fig.~\ref{fig:nikanika2}. The main objectives of LP-SZ encompass the in-depth study of ICM thermodynamic properties (pressure, density, temperature, mass profiles) and accurate measure of the mass-SZ observable scaling relation. It will be a key program to constrain the redshift evolution of the scaling relation and the pressure profile, as well as for testing the variation of cluster properties with morphology (departure from sphericity) or dynamical state (impact of mergers). This will have important implications for cosmology, such as improving the accuracy of cosmological constraints drawn from SZ-selected clusters.

\subsection{First SZ results and perspectives}

\begin{figure}
  \centering
  \includegraphics[width=0.49\textwidth]{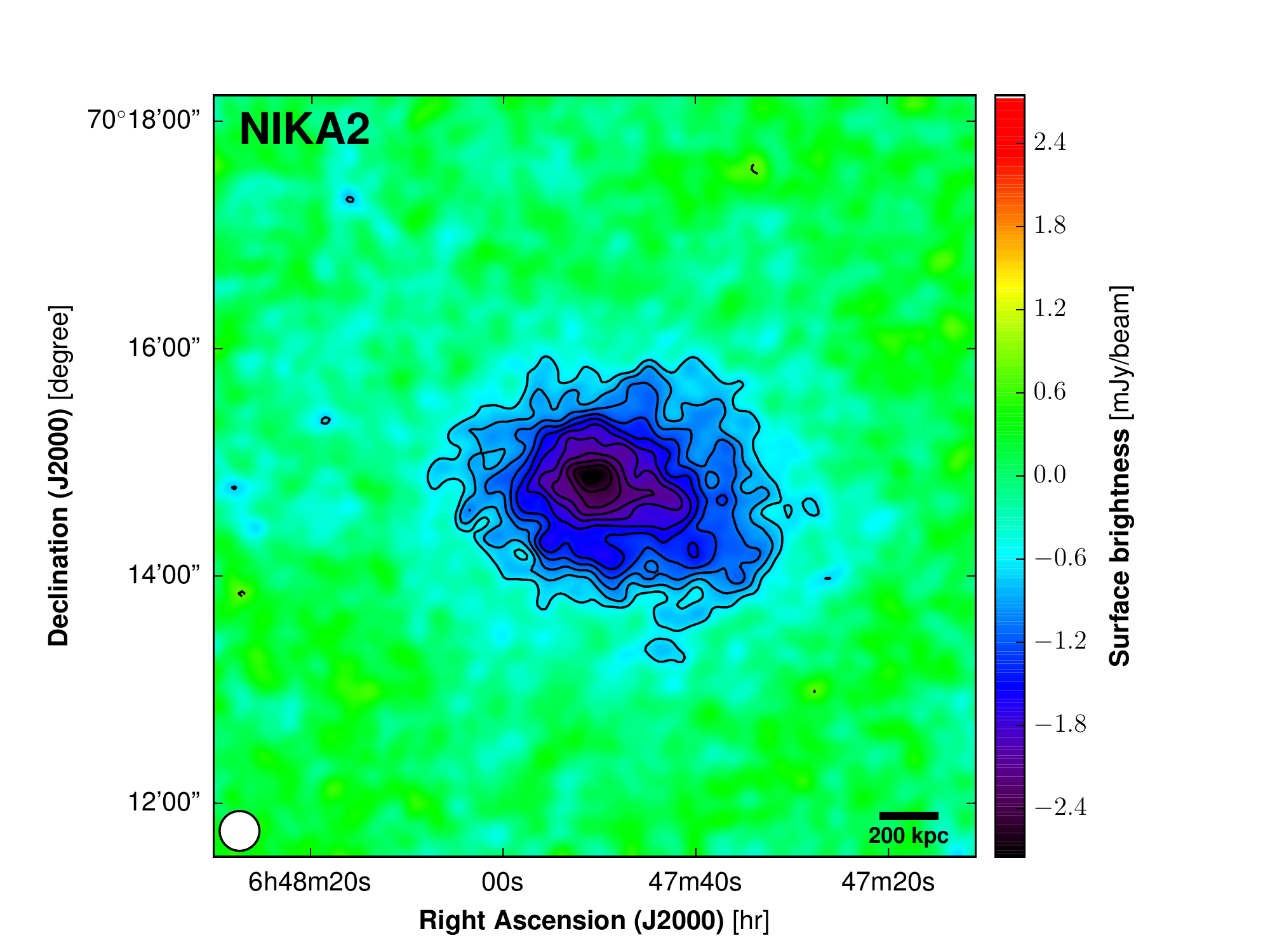}
  \includegraphics[width=0.49\textwidth]{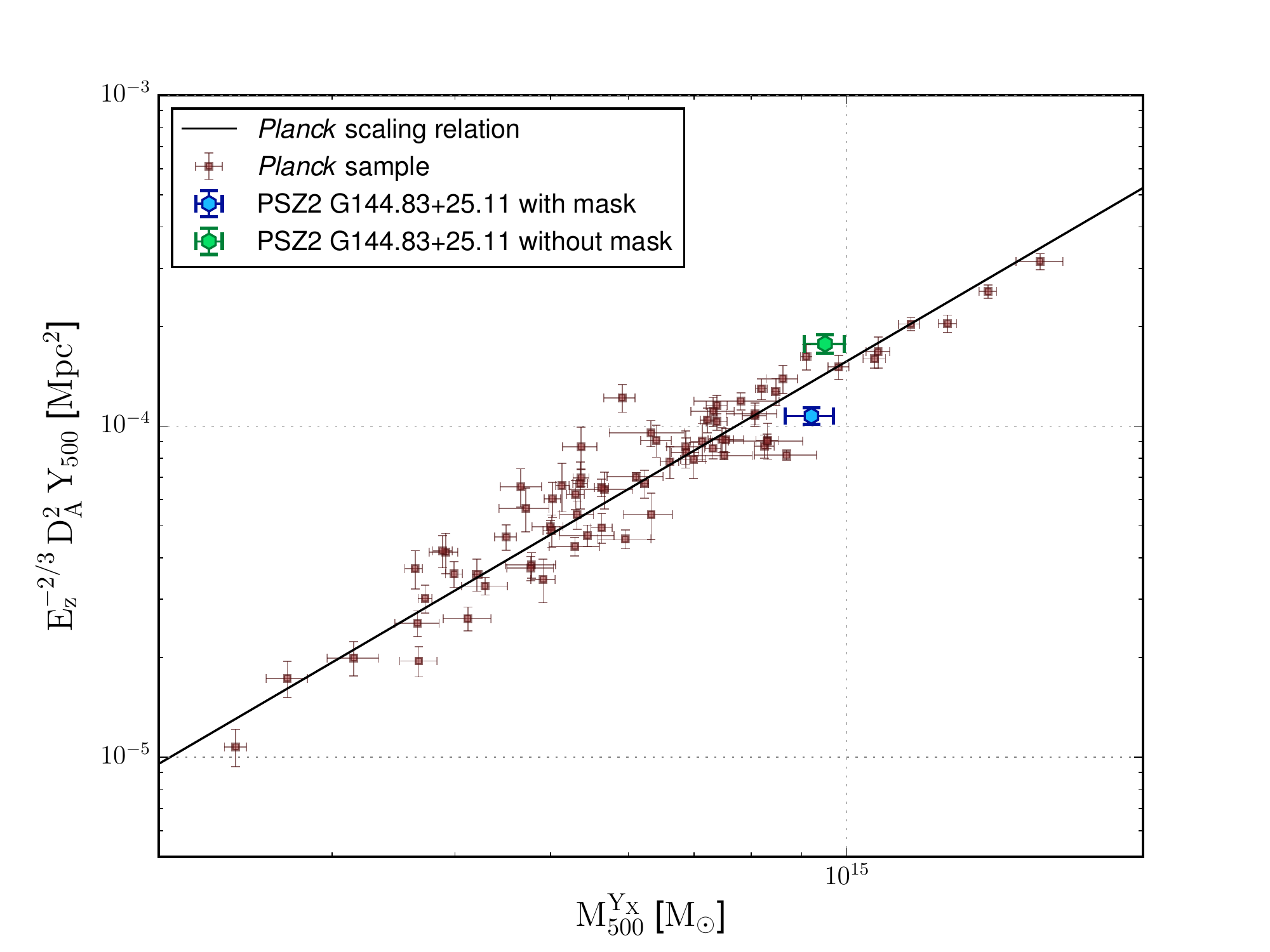}
  \caption{First SZ results with NIKA2. Left: The NIKA2 SZ map toward the galaxy cluster PSZ2-G0144.83+25.11. The high-resolution (20~arcsec) high-accuracy ($13.5\sigma $ measurement at peak) map covers the cluster from the core to the outskirts and reveals its morphology. An excess SZ signal is observed in the South-West region, indicating an overpressure within the intracluster medium (ICM). Right: Illustration of the impact of the ICM dynamics on the inner scatter of the SZ mass-observable relation. NIKA2 $Y_{500}$ estimates from the analysis with and without masking the over-pressure of PSZ2-G0144.83+25.11 are shown as a function of $M_{500}$, along with the cluster sample and the $Y_{500}-M_{500}$ scaling relation used in \emph{Planck} SZ-selected cluster count based cosmology analysis$^{13}$. These figures are extracted from Ruppin {\it et al.} (2018)$^{34}$. }
  \label{fig:nika2-sz}
\end{figure}

We report the first high-resolution thermal SZ mapping of a galaxy cluster with NIKA2~\cite{Ruppin2018}. As part of NIKA2 Science Verification Phase, the chosen target is PSZ2-G0144.83+25.11, a well-know cluster at intermediate redshift ($z=0.58$) and high mass ($M = 7.8 \times 10^{14} M_{\odot}$) from the second \emph{Planck} SZ catalog. The data set consists of 11 hours of integration time on source in poor weather conditions (opacity of 0.3 at 150GHz) during the April 2017 campaign.

We obtained a high-resolution ($20$~arcsec) Compton parameter ($y$) map toward PSZ2-G0144.83+25.11 with high-accuracy: we report a $13.5~\sigma$ measurement at peak and a detection up to 1.5 arcmin (about $600~\rm{kpc}$). NIKA2 $y$-map thus well covers this cluster from the core to the outskirts and reveals its morphology, as shown in the left panel of Fig.~\ref{fig:nika2-sz}.  Notably, we measure an excess of SZ signal in the South-West region, which indicates an over-pressure within the ICM.  
  
Other $y$-maps for this cluster are also available: the BOLOCAM~\cite{Sayers2013} arcminute angular resolution $y$-map that covers the cluster up to the outskirts and the MUSTANG~\cite{Young2014} high-resolution ($9$~arcsec) $y$-map of the center of the cluster, which both complement NIKA2 $y$-map.  

We combined NIKA2, BOLOCAM and MUSTANG $y$-maps and the integrated SZ signal from \emph{Planck} to measure the electronic pressure profile using a non-parametric deprojection~\cite{Ruppin:2016rnt}. When masking the over-pressure extension of the cluster in the combined $y$-map, we find a pressure profile compatible with the so-called universal pressure profile~\cite{A10}, which is derived from X-ray measurements, but not compatible with the pressure profile derived from the combination of BOLOCAM and MUSTANG (non-masked) $y$-maps~\cite{Young2014}. We report a significant discrepancy between the pressure profiles obtained with or without masking the over-pressure region. This discrepancy has in turn a sizable impact on $R_{500}$, the characteristic radius of the cluster, $Y_{500}$ the integrated Compton parameter up to $R_{500}$ and $M_{500}$, the mass enclosed within $R_{500}$, which are measured integrated quantity used for calibrating the mass-observable scaling law. In the right panel of Fig.~\ref{fig:nika2-sz}, NIKA2 $Y_{500}$ estimates for the cases with and without masking the over-pressure of PSZ2-G0144.83+25.11 are shown as a function of $M_{500}$, and compared to the $Y_{500}-M_{500}$ scaling relation used in \emph{Planck} SZ-selected cluster count based cosmology analysis~\cite{Ade:2015fva}. This illustrates the possible impact of the cluster ICM dynamics on the dispersion of the mass-observable scaling relation, the precise measure of which is required for accurate cosmology with clusters. NIKA2 will be a unique opportunity for a precise measurement of the mass-observable scaling relation for high-redshift clusters.

\section*{Acknowledgments}

We would like to thank the IRAM staff for their support during the campaigns. 
The NIKA dilution cryostat has been designed and built at the Institut N\'eel. 
In particular, we acknowledge the crucial contribution of the Cryogenics Group, and 
in particular Gregory Garde, Henri Rodenas, Jean Paul Leggeri, Philippe Camus. 
This work has been partially funded by the Foundation Nanoscience Grenoble, the LabEx FOCUS ANR-11-LABX-0013 and 
the ANR under the contracts "MKIDS", "NIKA" and ANR-15-CE31-0017. 
This work has benefited from the support of the European Research Council Advanced Grant ORISTARS 
under the European Union's Seventh Framework Programme (Grant Agreement no. 291294).
We acknowledge fundings from the ENIGMASS French LabEx (R. A. and F. R.), 
the CNES post-doctoral fellowship program (R. A.),  the CNES doctoral fellowship program (A. R.) and the FOCUS French LabEx doctoral fellowship program (A. R.). R.A. acknowledges support from Spanish Ministerio de Econom\'ia and Competitividad (MINECO) through grant number AYA2015-66211-C2-2.

\end{document}